\input{epsf}
\documentstyle[preprint,aps,amssymb]{revtex}
\draft

\begin{document}
\title{Dielectric response of charge induced correlated state in the
quasi-one-dimensional conductor (TMTTF)$_2$PF$_6$}
\author{F.~Nad$^{1,2}$, P.~Monceau$^1$, C.~Carcel$^3$, and J.M.~Fabre$^3$}
\address{$^1$Centre de Recherches sur les Tr\`es Basses
Temp\'eratures,\\
laboratoire associ\'e \`a l'Universit\'e Joseph Fourier, CNRS,
BP 166, 38042 Grenoble cedex 9, France\\
$^2$Institute of Radioengineering and Electronics,\\
Russian Academy of
Sciences, Mokhovaya 11, 103907 Moscow, Russia\\
$^3$Laboratoire de Chimie Structurale Organique,\\
Universit\'e de Montpellier, 34095 Montpellier cedex 5, France}
\maketitle

\begin{abstract}
Conductivity and permittivity of the quasi-one-dimensionsional 
organic transfer salt (TMTTF)$_2$PF$_6$ have been measured at low 
frequencies ($10^3-10^7$~Hz) between room temperature down to below 
the temperature of transition into the spin-Peierls state. We 
interpret the huge real part of the dielectric permittivity (up to 
10$^6$) in the localized state as the realization in this compound of 
a charge ordered state of Wigner crystal type due to long range 
Coulomb interaction.
\end{abstract}



\section{Introduction\protect\\}

Conductors formed of linear chains of organic mole\-cules
tetramethyltetrathiafulvalene (TMTTF) or
tetramethyltetraselenafulvalene (TMTSF) with a general formula
(TMTCF)$_2$X with C~= Se, S and X the interchain counterion~= ClO$_4$,
PF$_6$, Br, \ldots have been intensively studied these last years
because they exhibit a rich variety of cooperative phenomena
including superconductivity, antiferromagnetism (AF), spin density
wave (SDW), spin-Peierls state (SP), charge localization
\cite{JeromeSSC94}. While the Bechgaard (TMTSF)$_2$X salts
display a metallic behaviour down to low temperature where a
transition in a SDW state occurs below $\sim 12$~K, (TMTTF)$_2$X salts
exhibit a charge localization in the temperature range of 100--200~K,
with a maximum in conductivity at $T_\rho$ and a thermally activated
variation below $T_\rho$ \cite{CoulonJP82,LaversanneJPL84}, revealing
strong Coulomb interaction effects in these sulfur salts.

Quasi-one-dimensional conductor (TMTTF)$_2$X consists of
molecular chains (along the highest conductivity axis) with two
electrons per 4 molecules which corresponds to 1/4 filling in terms
of holes. These molecular chains are slightly dimerized due to
intermolecular interaction. As a result, with decreasing temperature
a dimerized gap $\Delta\rho$ opens with effective 1/2 filling
of the upper conduction band. Two intermolecular transfer integrals
along molecular stacks t$_1$ and t$_2$ have to be considered
\cite{FritschJPI91}. Quantum chemistry calculations show that
in (TMTTF)$_2$PF$_6$ the dimerization decreases with decreasing
temperature and at low temperatures t$_1$/t$_2$ ratio is
about 1.1 - 1.2 \cite{FritschJPI91}.

As was shown in several of theoretical and experimental works,
between many factors which determine the properties of
(TMTTF)$_2$X type salts, electron-electron correlation phenomena play
the leading part \cite{JeromeSSC94,FritschJPI91,EmeryPRL82,LeePRB77,%
PougetPRL76,%
SchulzIJMPB91,HirschPRB83,CaronPRB84,PencJPIV91,MilaPRB95,SeoJPSJ97}. In
this context two types of theoretical models have been essentially
developed. In the frame of the so-called g-ology models \cite{SchulzIJMPB91,%
CaronPRB84,BourbonnaisSM97} the electron-electron correlations are
considered as a perturbation to the one-electron approach. These models have
been used for describing the low energy properties of these salts which
exhibit the features of a Luttinger liquid rather
than a Fermi liquid \cite{BourbonnaisSM97}.

The second group of models includes the various versions of the Hubbard model
\cite{HirschPRB83,CaronPRB84,PencJPIV91,MilaPRB95,SeoJPSJ97,ClayPRB59}.
Extended Hubbard model takes
into account the interaction between charge carriers on the site of
host lattice (on-site interaction) with characteristic energy U as
well as the interaction between charge carriers on the neighboring
sites (near-neighbor interaction) with characteristic energy V.
In the case of (TMTTF)$_2$X compounds, at temperatures above the
transition into spin ordered states, the electron-electron interaction
is determined by long range Coulomb interaction and it is stronger
than spin interaction. This is one of the reason of spin-charge
separation observed in such 1D-conductors \cite{JeromeSSC94}. In
the frame of extended Hubbard model it was shown that the dimerized
energy gap is strengthened taking into account the on-site and
near-neighbor interactions \cite{HirschPRB83}. At the same time
the spectrum of spin exitations remains gapless which also
corresponds to spin-charge separation.

One important result of the extended Hubbard model approach
concerns the formation of a $4k_F$ CDW superlattice of
Wigner crystal type in such 1D compounds with decreasing temperature.
Using Monte Carlo technique it was shown that, for large enough U
and V magnitudes, strictly on-site interaction results only in a weak
$4k_F$ CDW, while long range Coulomb near-neighbor interaction
can produce a CDW singularity at $4k_F$ \cite{HirschPRB83}.
Analogously, using mean field approximation \cite{SeoJPSJ97}
it was recently shown that for one-dimensional molecular chain with
and without dimerization the form of the developed superstructure
depends considerably on the magnitude of near-neighbor interaction V:
at V above some critical value V$_c$ a $4k_F$ CDW superstructure
occurs with charge disproportionation depending on V.
Estimations of V/U and V/t$_2$ magnitudes in (TMTTF)$_2$PF$_6$ obtained
from quantum chemistry calculations and from optical conductivity
\cite{FritschJPI91,MilaPRB95} yields values for V/U in the range 0.4-0.5
and for V/t$_2$ in the range of 2-3 manifesting the essential role played
by the long range Coulomb interaction in this compound. Hereafter we
present results of conductivity and dielectric permittivity measurements
of (TMTTF)$_2$PF$_6$ which provide some evidence of a charge modulated
state (analogous to Wigner crystal) resulting from electron-electron
charge correlation below $T_\rho$.

\section{Experiment}

We have studied (TMTTF)$_2$PF$_6$ samples originating from two
batches. The crystals have been prepared using standard electrochemical
procedures \cite{DelhaesMCLC79}. Electrical contacts were prepared by
first evaporating gold pads on nearly the whole surface of the sample's ends
on which thin gold wires were attached afterwards with silver paste. We
have carried out the measurements of complex conductance $G(T,\omega)$,
using an impedance analyser HP~4192A in the frequency range
10$^3$--10$^7$~Hz and in the temperature range 4.2--295~K. The
amplitude of the ac voltage applied to the sample was within the
linear response and typically 30~mV/cm. We have noticed that the
cooling rate had a significant effect on the results of our
measurements, much more important than in the case of (TMTTF)$_2$Br
for instance \cite{NadEPJB98}: at a cooling rate above 0.5~K/mm,
cracks appear as seen in jumps in the temperature variation of $G$.
However with a slow cooling rate around 0.2~K/mm and along temperature
stabilisation before performing measurements, we succeeded in
recording the temperature dependences of real and imaginary parts of $G$
without any jumps for 3 samples. These samples have a length of 3--4~mm,
a cross-section about $2\times 10^{-5}-10^{-4}$~cm$^2$ and a room
temperature conductivity $\sim 40~\Omega^{-1}$cm$^{-1}$. The results
obtained for these 3 samples being qualitatively similar and we will
present the data for two of them (referred as samples 1 and 2).

\begin{figure}[t]
\centerline{\epsfxsize=12cm \epsfbox{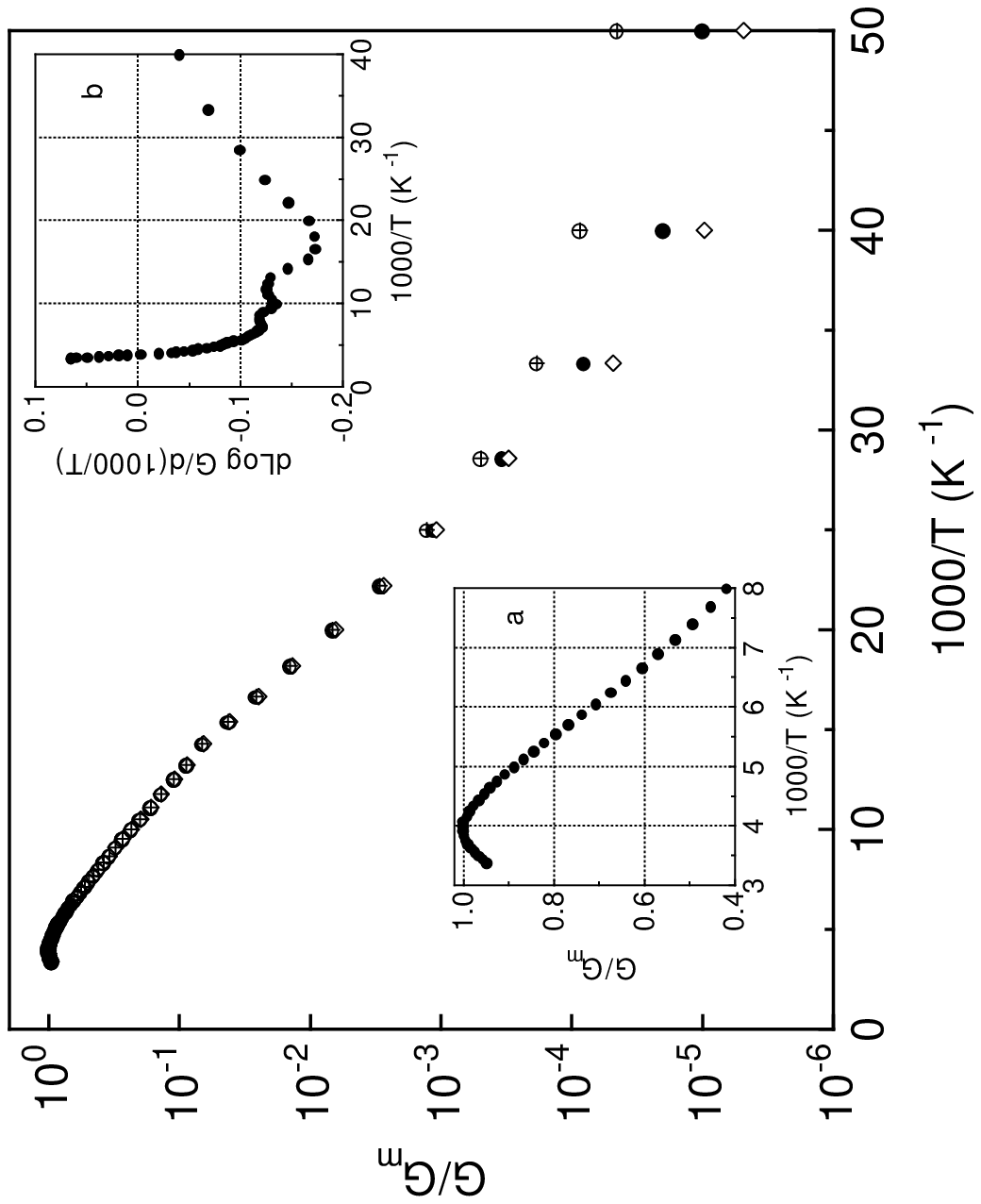}}
\caption{Variation of the real part of the conductance $G$ (sample~1)
normalized to its maximum $G_m$ as a function of the inverse
temperature at frequencies (in~kHz): $\lozenge$~10,
{\Large $\bullet$}~100,
$\oplus$~1000. Inset~(a): details of the temperature dependence
of $G/G_m$ near the maximum. Inset~(b): temperature dependence of the
logarithmic derivative $d\log G/d(1000/T)$.}
\label{fig1}
\end{figure}

Fig.\ref{fig1} shows the variation of the conductance $G(T)$
of sample~1 normalized by its maximum value $G_m$ as a function of the
inverse temperature. The detailed $G(T)$ dependence near the room
temperature is shown in inset~(a) of Fig.\ref{fig1}. With
decreasing temperature, the conductance of (TMTTF)$_2$PF$_6$ first
grows up to a maximum at $T_\rho$~= 250~K as previously reported in
\cite{CoulonJP82,LaversanneJPL84}. Below $T_\rho$, the decrease of
$G(T)$ in the temperature range 200--70~K follows an Arrhenius type
behavior with an activation energy $\Delta\rho\simeq 300$~K. The value
of this activation energy is in good agreement with the evaluation of
charge gap in \cite{PencJPIV91}. It was shown
that $\Delta\rho=1/4(t_1+t_2)$ which provide $\Delta\rho\simeq 300$~K
for (TMTTF)$_2$PF$_6$.
At the same time this magnitude of energy gap is two times smaller
than the value reported in a previous publication \cite{CoulonJP82}.
This difference can be a result of the gap determination  in
\cite{CoulonJP82} from $G(1/T)$ dependence obtained by averaging of
$G(1/T)$ dependencies from several different samples with jumps of $G$
due to cracks.The jump-like decrease of conduction at every crack
results in more steep averaged $G(1/T)$ dependence and accordingly
in more higher gap magnitude. The small cooling rate enables us to
avoid cracks and to determine more accurately the energy gap
magnitude which agrees with theoretical evaluation \cite{PencJPIV91}.

At lower temperature we observed a bend on the $G(1/T)$ dependence near
70~K, i.e. $G$ begins to decrease more faster (activation energy 
$\approx$~380~K) with a following gradual
transition to a new activation regime at temperatures below 25~K. Detailed
measurements of sample 2 show that the decrease of $G$ is thermally
activated between 4.2~K and 20~K with an activation energy $\simeq 36$~K.
This energy gap satisfied the BCS-like relation $\Delta=1.75kT_{\rm SP}$
with $T_{\rm SP} \simeq 20$~K as previously shown in the temperature
dependence of the EPR spin susceptibility and the nuclear relaxation
rate $T_1^{-1}$ \cite{WzietekJPI93}.

It can also be seen in Fig.\ref{fig1} that the frequency
dependence of $G(1/T)$ become noticeable below 70~K with a frequency
dispersion growing at lower temperature. For showing more precisely
the particular points on the $G(1/T)$ variation we draw in inset~(b)
of Fig.\ref{fig1} the temperature variation of the logarithmic
derivative which was reproducible for our three samples.
The decrease of $G$ near $T\simeq 70$~K may indicate the beginning 
of a transition into some new ground state.
Two additional small minima in $d\log G/d(1/T)$ are also visible
near 140~K and 100~K (inset~(b) in Fig.\ref{fig1}.

\begin{figure}
\centerline{\epsfxsize=14.5cm \epsfbox{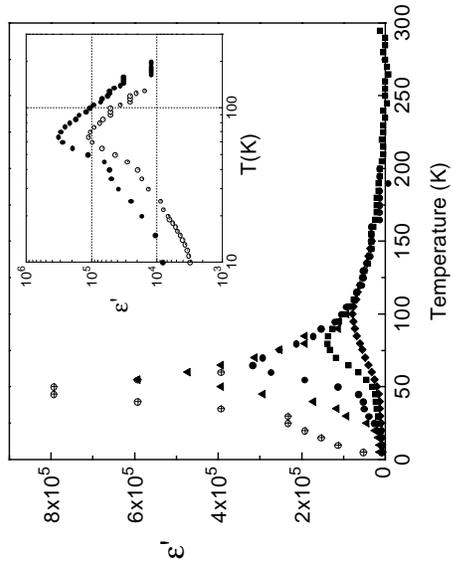}}
\caption{Temperature dependence of the real part of the dielectric
permittivity $\epsilon^\prime$ (sample~1) at frequencies (in~kHz):
$\oplus$~1, $\blacktriangle$~10, {\Large $\bullet$}~100,
$\blacksquare$~1000,
$\blacklozenge$~5000. Inset: temperature dependence of
$\epsilon^\prime$ for sample~1 ({\Large $\bullet$}) and sample~2~($\odot$) at
100~kHz in a double logarithmic scale.}
\label{fig2}
\end{figure}

The temperature variation of the real part of the dielectric
permittivity $\epsilon^\prime(T)$ of sample~1 in the frequency range
$10^3-5\times 10^6$~Hz is shown in Fig.\ref{fig2}.
Between 300~K and 220~K, the magnitude of $\epsilon^\prime$ is below
the background level determined by the resolution of our measurements
in this temperature range.  The growth of $\epsilon^\prime$ is
noticeable below $\simeq 200$~K and frequency independent down to
$T\simeq 110$~K. Below this temperature a significant frequency
dispersion occurs as seen in Fig.\ref{fig2}: at a given frequency
$\epsilon^\prime(T)$ goes through a maximum before falling down. With
decreasing frequency, the amplitude of the maximum of
$\epsilon^\prime(T)$ is larger and the maximum position on the
temperature scale shifts to lower temperature. This behavior is
qualitatively similar to critical slowing down phenomena near a phase
transition. The $\epsilon^\prime(T)$ curves for samples 1 and 2 are
shown in inset of Fig.\ref{fig2} in a double logarithmic scale,
manifesting their qualitative similarity. In the temperature range
200--70~K, the $\epsilon^\prime(T)$ dependence can be described by
the power law: $\epsilon^\prime(T)\sim T^{-\alpha}$ with
$\alpha\simeq 1/3$. While $\epsilon^\prime(T)$
is decreasing from 70~K down 10~K, a small bump in
$\epsilon^\prime(T)$ can be seen near $\simeq 35$~K.

\begin{figure}
\centerline{\epsfxsize=14.5cm \epsfbox{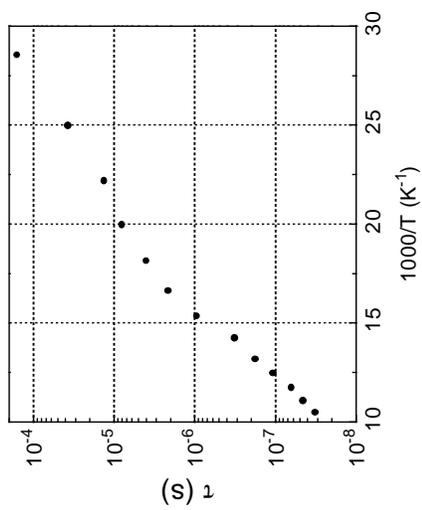}}
\caption{Temperature dependence of the relaxation time $\tau$ of the
dielectric relaxation (sample~1).}
\label{fig3}
\end{figure}

The frequency dependencies of the conductance, of the real
part $\epsilon^\prime$ and of the imaginary part
$\epsilon^{\prime\prime}$ of the dielectric permittivity of
(TMTTF)$_2$PF$_6$ have a form similar to those we previously measured
on other 1D organic compounds \cite{NadEPJB98,NadSSC95}. As usual, the
frequency, $f_m$, corresponding to the maximum of
$\epsilon^{\prime\prime}(T)$
corresponds to some mean value of the relaxation time $\tau=1/2\pi
f_m$ of charge polarization. The variation of $\tau$ with the inverse
temperature is drawn in Fig.\ref{fig3}
for sample~1 in the temperature rangee 95--35~K: for $60<T<95$~K,
$\tau(1/T)$ is thermally activated with an energy activation of $\sim
650$~K; but, at lower $T$, 35~K$<T<$60~K, the activation energy is 
smaller, $\approx$~380~K, the same as the activation energy of the 
conductivity in same temperature range.

\section{Discussion}

As follows from published data \cite{JeromeSSC94,WzietekJPI93}
and from our experimental results, (TMTTF)$_2$PF$_6$ can be
characterized by two distinct energy scales: temperature
$T_\rho$~= 250~K corresponding to the conduction maximum and
$T_{\rm SP} \simeq 20$~K corresponding to the transition into
the spin-Peierls state. If one will consider the temperatures
$T_{\rm SP}<T<T_\rho$ as a range of simple localization of
charge carriers \cite{JeromeSSC94}, one could try to explain
the observed growth of $\epsilon^\prime$ as a result of the growth
of $2k_F$ CDW fluctuations when approaching $T_{\rm SP}$
\cite{PougetJP96}. As was mentioned in \cite{SchulzIJMPB91}
the temperature range of fluctuations near the spin-Peierls
transition can be wide enough and reach $\sim 3T_{\rm SP}$.
It means that in (TMTTF)$_2$PF$_6$ the manifestation of
the fluctuations can be noticed beginning from
$T=T_{\rm SP} + 3 \times T_{\rm SP} \simeq 80$~K, temperature at
which $2k_F$ fluctuations have started to be really observed
\cite{PougetJP96}. However as can be seen from Fig.\ref{fig2}
the $\epsilon^\prime$ growth begins nevertheless not from 80~K
but from $\simeq 200$~K. For frequencies above 10$^6$~Hz this
$\epsilon^\prime$ growth is rather achieved at 80~K. This
dismatching in the temperature ranges for occurence of $2k_F$
fluctuations and the $\epsilon^\prime$ growth makes difficult
the explanation of $\epsilon^\prime$ growth as a result of
$2k_F$ CDW fluctuations. In the same temperature range
50 -- 200~K, EPR susceptibility $\chi$ decreases monotonously
without any maximum \cite{WzietekJPI93}.A small maximum on
$\chi(T)$ dependence was observed near 40~K with a following
decrease of $\chi$ as a result of the transition to the spin-Peierls
state. Such considerable qualitative difference between
$\epsilon^\prime(T)$ and $\chi(T)$ dependences confirms
the existence of spin-charge separation in (TMTTF)$_2$PF$_6$ salts.

We tentatively ascribe the frequency and temperature dependences
of $\epsilon^\prime$ of (TMTTF)$_2$PF$_6$ in the temperature range
$T_{\rm SP}<T<T_\rho$ as related to the charge induced correlation
phenomena discussed above. Taking into account only on-site interactions
in the Hubbard model for half-filled band, the commensurate charge 
induced superstructure ($4k_F$ CDW) is strictly linked to the host
lattice. In such a case the possibility of polarization of
the superstructure, i.e. its shift with respect to the host lattice,
is small and consequently one would expect a low magnitude of
the dielectric permittivity. Ground state with Mott-Hubbard gap and
appropriate charge localization have been realized in 3D semiconductors
\cite{Mott74}. In such compounds $\epsilon^\prime$ is in the order of
10, a typical value for usual semiconductors, while in our samples of
(TMTTF)$_2$PF$_6$ the $\epsilon^\prime$ magnitudes amount by several
orders of value larger (Fig.\ref{fig2}).

As explained above, long range Coulomb interaction of sufficient
strength and appropriate charge induced correlation may lead to
a superstructure with charge disproportionation corresponding to a $4k_F$
CDW as in a Wigner crystal. For (TMTTF)$_2$PF$_6$ this charge
disproportionation can be evaluated on the base of calculations in
\cite{SeoJPSJ97} taking into account the estimated magnitude of
the reduced near-neighbor Coulomb interaction $V/t_2 \simeq 2$ \cite
{FritschJPI91,MilaPRB95}.For such a $V/t_2$ value we estimate the charge
disproportionation as about 1:3. Such a degree of disproportionation and
the possibility of its variation with temperature \cite{FritschJPI91},
provide some evidence that this charge supertructure is probably more soft,
more weakly connected to the host lattice and consequently more easily
polarizable than in the case of only on-site interaction. We ascribe
the large magnitude of the dielectric permittivity, which we have found
out below $T_\rho$, to the collective response of such charge superstructure
of Wigner crystal type with charge disproportionation formed
in (TMTTF)$_2$PF$_6$. Indeed, as can be seen from Fig.\ref{fig2} in the
temperature range above $T_\rho$ the $\epsilon^\prime$ magnitude does not
exceed the background level. Its noticable growth begins below 200~K when,
as we considered, the growing of this Wigner type CDW superstructure
begins to determine the kinetic properties of the compound.
The $\epsilon^\prime$ growth with temperature decreasing is probably
associated with the gradual enhancement of the CDW superstructure. Possible
reasons for such an enhancement can be the growth of intrachain charge induced
correlations as well as the growth of interchain interactions. Both of
them favour the three dimensional ordering of the CDW superstructure,
i.e. the formation of a 3D electronic crystal. The maximum value of
$\epsilon^\prime$ amounts to 10$^5$ -- 10$^6$, 2 or 3 orders of magnitude
lower than the $\epsilon^\prime$ values in incommensurate charge
\cite{NadPR} and spin \cite{NadSSC95} density wave below their
transition temperature. However the magnitude of $\epsilon^\prime$ in
(TMTTF)$_2$PF$_6$ is nearly comparable with that in (TMTTF)$_2$Br
\cite{NadEPJB98}.

According to X-ray measurements \cite{PougetJP96}, diffuse
$2k_F$ scattering grows critically in (TMTTF)$_2$PF$_6$ below $\sim
80$~K which originates from the gradual enhancement of spin induced
electron-electron correlation of antiferromagnetic type. Due to the
growth of the electron-phonon interaction and of the increasing of
the interchain interaction, these $2k_F$ spin induced correlations
diverge below $T_{\rm SP}$ resulting in the condensation of an
ordered spin-Peierls state with a $2k_F$ superstructure.

The opening of the spin-Peierls energy gap and the two-fold commensurability
of the superlattice lead to the freezing of charge polarization degrees of freedom
and consequently to the sharp decrease of $\epsilon^\prime$. The slowing down
behavior of $\epsilon^\prime$ and the temperature dependence of the relaxation
time (Fig.3) also indicate the lattice involvment (i.e. heavy molecules)
in the relaxation process.

\section{Conclusion}

In conclusion, our measurements of the complex conductivity of 
(TMTTF)$_2$PF$_6$ show the main following
features: 1)~development of a charge energy gap the magnitude of which
corresponds to theoretical evaluations in the frame of extended Hubbard
model; 2)~existence of peculiarities on $G(1/T)$ dependence, for example,
a minimum of the logarithmic derivative near 60~K; 3)~in the same temperature
range, finding of a huge maximum of the real part of the dielectric permittivity
(up to 10$^6$) with a slowing down behavior with decreasing temperature
while the magnetic susceptibility does not show any significant
variation, which corresponds to large charge
polarization simultaneously with spin-charge separation; 4)~considerable
difference between the temperature dependence of the dielectric permittivity and
$2k_F$ diffuse X-ray scattering.

All these features seem to confirm
the possibility of the formation in the temperature range
$T_{\rm SP}<T<T_\rho$ of a charge ordered state with a high
polarizability. On the basis of our experimental results and some theoretical
approaches we consider that we consider that the huge amplitude of 
the real part of the dielectric permittivity of (TMTTF)$_2$PF$_6$ can 
hardly be provided by a Mott-insulator. On the contrary, we argue that 
this large dielectric polarizability reflects the collective response 
of $4k_F$ charge density wave of Wigner crystal type due to long range
Coulomb interaction and electron-electron correlation.
Recently charge disproportionation \cite{HirakiPRL98} and
$4k_F$ superlattice \cite{NogamiSM99} have been reported from NMR and
X-ray measurements in a 1/4 filled one-dimensional organic compound
(DI-DCNQI)$_2$Ag without dimerization.

{\bf Acknowledgments}

We would like to thank S.~Brazovskii and N.~Kirova for helpful
discussions, and D.~Staresinic for help in the experiment. Part of
this work was supported by the Russian Fund for Fundamental Research
(grant N$^\circ$~99-02-17364) and the twinning research programme
N$^\circ$~19 (grant N$^\circ$~98-02-22061) between CRTBT-CNRS and
IRE-RAS.

\newpage


\end{document}